\documentclass[10.75pt,preprint]{article}
\usepackage[english]{babel}
\usepackage{graphicx}
\usepackage{amsmath}
\usepackage{amsfonts}
\usepackage{amssymb}

\begin{document}

\title{VELOCITY AND DISTRIBUTION OF PRIMORDIAL NEUTRINOS.}
\author{Jorge Alfaro and Pablo Gonz\'alez.\\ \textit{Facultad de F\'isica, Pontificia Universidad Cat\'olica de Chile.}\\
\textit{Casilla 306, Santiago 22, Chile.}\\ jalfaro@uc.cl,
pegonza2@uc.cl}

\maketitle

\section*{ABSTRACT}
The Cosmic Neutrinos Background (\textbf{CNB}) are Primordial
Neutrinos decoupled when the Universe was very young. Its
detection is complicated, especially if we take into account neutrino mass
and a possible breaking of  Lorentz Invariance at high energy, but has a
fundamental relevance to study the Big-Bang. In this paper, we
will see that a Lorentz Violation does not produce important
modification, but the mass does. We will show how the neutrinos
current velocity, with respect to comobile system to Universe
expansion, is of the order of $1065$ $\left[\frac{km}{s}\right]$, much less than
light velocity. Besides, we will see that the neutrinos
distribution is complex due to Planetary motion. This
prediction differs totally from the usual massless case, where we
would get a correction similar to the Dipolar Moment of the
\textbf{CMB}.\\

\section*{INTRODUCTION}

From the beginning, the photons and all particles were coupled
forming a plasma that was evolving under the influence of the
Universe expansion. In such a moment, when the photons were
dominating the expansion, the neutrinos were decoupling from the
plasma and evolved in an independent way. One of the last
discoveries about neutrino is its mass. This has relevant effects
in the Standard Model and in some of its characteristics,
distinguishing it from the photons. One of them, which we will
study, is its velocity. Thus, we will analyze the evolution of the
neutrinos's kinetic energy since its decoupling till today.\\

Other phenomenon that we will study, that is directly related with
the first one, is the neutrinos distribution. The detection of the
Cosmic Microwave Background of Photons (\textbf{CMB}) is the best
proof of the Big-Bang scenario \cite{abg} that helped to check or
refute models that describe it, and study the composition of the
Universe. Because of this, it is important to study the Cosmic
Neutrinos Background (\textbf{CNB}), especially the form of their
Distribution Function to consider the effect of the peculiar
velocity of the planet, named Dipolar Moment in the \textbf{CMB},
and optimize the detection. This is already complicated due to the
low interaction that the neutrinos have with ordinary matter. The
calculation will be done for photons and neutrinos in parallel.\\

 Finally, we will include a Lorentz Invariance Violation (LIV)
represented by an alteration to the Dispersion Relation of energy
given by \cite{liv,Alfaro1,Alfaro2,Alfaro3}:

$$E^2 = v_{max}^2p^2 + m^2c^4$$

Where $v_{max} = c(1-\alpha)$ is the maximum attainable particle velocity
with $\alpha \sim (10^{-22}-10^{-23})$ . The motivation to use
this LIV comes from the possibility that, at the  high energies
available in the Big Bang there take place some LIV due to Quantum
Gravity \cite{liv,Alfaro2,Alfaro3}. If such a LIV exists, the first problem is
the appearance of a privileged reference system, but fortunately
exists a natural candidate, the one where the \textbf{CMB} is isotropic. 
A LIV without a preferred frame as in Double Special Relativity \cite{dsr}, will not be considered here. \\


\section{ENERGY AND  VELOCITY OF PRIMORDIAL NEUTRINOS.}

Initially the neutrinos were in thermal equilibrium with the rest
of matter. For this, is necessary that $\Gamma_{i} \gg H$, where
$\Gamma_{i}$ is the rate of interactions of the species $i$, $H
\propto T^2$ is Hubble's constant and $T$ the temperature. While
the neutrinos are kept in equilibrium, its distribution will be
given by Fermi-Dirac's statistics:

$$f_{eq}(E,T) = \frac{1}{e^{\frac{E-\mu}{k_BT}}+1}$$

During cosmic expansion, the temperature will be diminishing down
to a point where $\Gamma_{\nu} \lesssim H$ and $\Gamma_{i \neq
\nu} \gg H$. This means that the neutrinos lost the equilibrium and
are decoupled from the rest of the matter. We will name
$T_{\nu,D}$ the neutrinos decoupling temperature that are obtained
when we impose $\Gamma_{\nu} \simeq H(T) $. To see what is
happening with its distribution we will do the following analysis.
For a time $t_0$, an observer sees in any direction a quantity $dN
= f d^3r d^3p$ of neutrinos in an volume $d^3r$ and with momentum
between $\vec{p}$ and $\vec{p} +d\vec{p}$. After a $dt$ time, the
neutrinos have not interacted, so $dN$ remains constant, but the
volume in which they are, have increased in a factor
$\left(\frac{R(t_0+dt)}{R(t_0)} \right)^3$ and the momentum has
diminished in $\frac{R(t_0)}{R(t_0+dt)}$, because of the expansion
of the Universe. This means that $f(E,T_{\nu}) $ is constant in
time. Therefore, for $t> t_D$ (or $T_{\nu} < T_{\nu,D}$) with
$t_D$ the moment in which is produced the decoupling, the
distribution function is given by \cite{Early uni}:

\begin{equation}
\label{feq} f[E(p(t)),T_{\nu}(t)] = f_{eq}[E(p_D),T_{\nu,D}] =
f_{eq}\left[E\left(p(t)\frac{R(t)}{R_D}\right),T_{\nu,D}\right]
\end{equation}

When the subscript $D$ refers to the age of decoupling. In
addition, we know that the number of neutrinos, the total energy
and the energy per neutrino are given by:

\begin{equation}
\label{Nf} N_{\nu} = \frac{gV}{(2\pi\hbar)^3} \int
f(p,T_{\nu})d^3p
\end{equation}

\begin{equation}
\label{Ef} E_{\nu} = \frac{gV}{(2\pi\hbar)^3} \int E(p)
f(p,T_{\nu})d^3p
\end{equation}

\begin{equation}
\label{ef} \varepsilon_{\nu} = \frac{E_{\nu}}{N_{\nu}}
\end{equation}

Where:

$$E^2(p) = v_{max}^2 p^2 + m^2c^4$$

to allow for a small LIV in the dispersion relation.\\

Now we can determine the Distribution Function that they will have
after being decoupled. It is possible to express the energy of the
neutrinos (high energies and small masses) as $E(t) = v_{max,\nu}
p(t)$ during the decoupling (We use an expansion with zero order
in the mass because $f$ depends exponentially on $E$), and as $p_D
= p(t) \frac{R(t)}{R_D}$ we obtain using (\ref{feq}):

\begin{equation}
\label{f_Rel} f[p,T_{\nu}] =
\frac{1}{e^{\frac{v_{max,\nu}p}{k_BT_{\nu}}}+1}
\end{equation}

With $T _{\nu} = T _{\nu, D} \frac{R_D}{R (t)}$ and $\mu _{\nu}
=0$ because of the low interaction that they have with matter.
This means that the distribution of neutrinos after decoupling is
Fermi's with temperature $T_{\nu}$, therefore $RT_{\nu} = cte$.
Replacing it in (\ref{Nf}) and (\ref{Ef}):

$$N_{\nu} = \frac{gV}{(2\pi\hbar)^3} \int \frac{1}{e^{\frac{v_{max,\nu}p}{k_BT_{\nu}}}+1} d^3p$$
$$E_{\nu} = \frac{gV}{(2\pi\hbar)^3} \int \frac{E(p)}{e^{\frac{v_{max,\nu}p}{k_BT_{\nu}}}+1} d^3p$$

Naturally, $N_{\nu}$ will be constant in time. Using the change of
variable $x = \frac{v _{max,\nu} p}{k_BT _{\nu}}$, we obtain:

\begin{equation}
\label{N} N_{\nu} =
\frac{3gV\zeta(3)(k_BT_{\nu})^3}{4\pi^2\hbar^3v_{max,\nu}^3}
\end{equation}

Where $\zeta (3) = 1.2021$ is the Riemann's Zeta function. We see
that, in fact, $N_{\nu}$ keeps constant in time because $V \propto
R^{3} (t) $ and $T_{\nu} \propto R^{-1}(t) $. To determine
$E_{\nu}$, we must compute the integral, which is complicated for
the general case. Thus , we will analyze the extreme cases where
the neutrinos continue being relativistic and when they do not.
Due to the spherical symmetry, the velocity is only radial, therefore
we just must determine its modulus. The modulus of the velocity of
a particle is given by:

$$v = \frac{\partial \varepsilon}{\partial p}$$

Being $\varepsilon$ and $p$ the energy and the momentum of a
particle, related by our dispersion relation:

$$\varepsilon^2 = v_{max}^2 p^2 + m^2c^4$$

While the particle continues being relativistic, developing the
derivative till the second order in the mass ($\varepsilon \gg
mc^2$), we obtain:

\begin{equation}
\label{vrel} v_{\nu} \simeq
v_{max,\nu}\left(1-\frac{1}{2}\left(\frac{mc^2}{\varepsilon}\right)^2\right)
\end{equation}

Notice that we must use $E (p) = v_{max} p$ to calculate
$E_{\nu}$, to the order of approximation in the mass that we are
considering.\\

Now, if the particle becomes Non-Relativistic, we have that the
energy and the velocity of a particle to second order in the
momentum ($pv_{max, \nu} \ll mc^2$) will be:

$$\varepsilon_{\nu} \simeq mc^2 + \left(\frac{v_{max,\nu}}{c}\right)^2\frac{p^2}{2m}$$

\begin{equation}
\label{vnorel} v_{\nu} = \frac{\partial \varepsilon}{\partial p}
\simeq \left(\frac{v_{max,\nu}}{c}\right)^2\frac{p}{m} =
v_{max,\nu}\sqrt{2\left(\frac{\varepsilon}{mc^2}-1\right)}
\end{equation}

Where we see that, to keep the order in the momentum, the
calculation must be up to second order in the expression of
$E(p)$, therefore $E(p) = mc^2 +
\left(\frac{v_{max,\nu}}{c}\right)^2\frac{p^2}{2m}$.\\

\subsection{Relativistic Neutrinos} As we said, to determine
$E _{\nu}$ we must use $E (p) = v _{max} p$. With this, we obtain
the expression:\\

\begin{equation}
\label{Erelaun} E_{\nu} =
\frac{7\pi^2gV(k_BT_{\nu})^4}{240\hbar^3v_{max,\nu}^3}
\end{equation}

Using (\ref{N}) and (\ref{Erelaun}) in (\ref{ef}) and
(\ref{vrel}), we obtain:

\begin{equation}
\label{ener_rel} \varepsilon_{\nu} =
\frac{7\pi^4}{180\zeta(3)}k_BT_{\nu}
\end{equation}

\begin{equation}
\label{veldesrel_rel} v_{\nu} =
v_{max,\nu}\left(1-\frac{1}{2}\left(\frac{180\zeta(3)m_{\nu}c^2}{7\pi^4k_BT_{\nu}}\right)^2\right)
\end{equation}

If we define the relative velocity between the neutrinos and the
photons as $\Delta v = c - v_{\nu}$, result:

$$\Delta v = \Delta
v_{max}+\frac{v_{max,\nu}}{2}\left(\frac{180\zeta(3)m_{\nu}c^2}{7\pi^4k_BT_{\nu}}\right)^2$$

Where $\Delta v_{max} = c - v_{max, \nu} = c\alpha_{\nu}$. We can
see that this factor vanishes if the violation does not exist.
Evaluating numerically:

\begin{equation}
\label{c-veldesrel_rel} \frac{\Delta v}{c} =
\alpha_{\nu}\left(1-5.04 \times
10^{-2}\left(\frac{M_{\nu}}{k_BT_{\nu}}\right)^2\right)+5.04
\times 10^{-2}\left(\frac{M_{\nu}}{k_BT_{\nu}}\right)^2
\end{equation}

Where we have separated the LIV dependent part from the rest.\\

\subsection{Non-Relativistic Neutrinos} In this case we have that
$E (p) = m _{\nu} c^2 + \left (\frac{v _{max, \nu}}{c} \right)
^2\frac{p^2}{2m _{\nu}}$, therefore, when we evaluate in $E_{\nu}$
using (\ref{N}), we obtain:\\

\begin{equation}
\label{Eyanorel}E_{\nu} = N_{\nu}m_{\nu}c^2 \left(1 +
\frac{1}{2}\left(\frac{k_BT_{\nu}}{m_{\nu}c^2}\right)^2
\frac{I_4}{I_2}\right)
\end{equation}

With $I_n = \int_0^{\infty} \frac{x^n}{e^x +1} dx = \left
(1-\frac{1}{2^n} \right) n! \zeta (n+1) $. Then, evaluating in
(\ref{ef}) and (\ref{vnorel}), we have:

\begin{equation}
\label{eneryanorel}
\varepsilon_{\nu} = m_{\nu}c^2 \left(1 +
15\frac{\zeta(5)}{2\zeta(3)}\left(\frac{k_BT_{\nu}}{m_{\nu}c^2}\right)^2
\right)
\end{equation}

\begin{equation}
\label{veldesrel_norel} v_{\nu} =
v_{max,\nu}\sqrt{15\frac{\zeta(5)}{\zeta(3)}}\frac{k_BT_{\nu}}{M_{\nu}}
\end{equation}

Giving a relative velocity:

\begin{equation}
\label{c-veldesrel_norel}\frac{\Delta v}{c} =
\alpha_{\nu}\sqrt{15\frac{\zeta(5)}{\zeta(3)}}\frac{k_BT_{\nu}}{M_{\nu}}
+
\left(1-\sqrt{15\frac{\zeta(5)}{\zeta(3)}}\frac{k_BT_{\nu}}{M_{\nu}}\right)
\end{equation}

Where we have separated the LIV part from the rest and $\zeta (5)
= 1.0369$. Since the neutrino velocity cannot be higher than its
maximum velocity, we must see to what temperatures this
approximation is valid. We have that $v_{\nu}> v_{max, \nu}$ if
$k_BT_{\nu}> \sqrt{\frac{\zeta(3)}{15\zeta(5)}} M_{\nu}$. It means
that the approximation is valid if $k_BT_{\nu} \ll \sqrt
{\frac{\zeta (3)}{15\zeta(5)}} M_{\nu} \sim 0.28 M_{\nu}$.\\

\subsection{Numerical Results and Analysis}

In the age of decoupling of the neutrinos, we know that
$k_BT_{\nu,D} \simeq (2 - 4)$ [MeV] and, currently, $k_BT_{\nu, 0}
= 1.68 \times 10^{-4}$ [eV]. In addition to this, for cosmological
parameters, we know \cite{masa}:

$$\sum_{i} m_{\nu_i} \leq 0.17 [eV]$$

That clearly indicates that they are relativistic in the moment of
the decoupling. There exist many estimations of the masses of the
neutrinos, but none of them are very precise. Thus, we will use
$m_{\nu} \simeq 0.17 [eV]$. This way, we are sure of being inside
the correct limits and we will find the maximum effect that the
mass could have in the velocity of neutrinos. This way, none of
these estimations is below $k_BT_{\nu,0}$, therefore they are
Non-Relativistic nowadays.\\

Before discussing the results, we will analyze the effect of the
LIV. Thus, we will compare our relativistic expressions with our
Non-Relativistic ones. If we observe these expressions, we can see
that both are proportional to $v_{max, \nu}$, which is the only
thing that depends on $\alpha_{\nu}$. It means that the difference
in percentage between the case with and without LIV is always:

$$\frac{v_{\nu}(0) - v_{\nu}(\alpha_{\nu})}{v_{\nu}(0)}100\% = \alpha_{\nu} 100\% = 1 \times 10^{-20} \%$$

Therefore, it is not possible that this LIV has an important
effect in the neutrinos, then we will continue our calculations
using $\alpha=0$.\\

Previously we mentioned that our  Non-Relativistic approximation is
valid if $\frac{k_BT_{\nu}}{M_{\nu}} \leq 0.28$. At present we
have that $\frac{k_BT_{\nu}}{M_{\nu}} \simeq 10^{-3}$ fulfilling
the Non-Relativistic bound, but with a mass $100$ times minor
($\sim 2 \times 10^{-3}$ [eV]) the bound is not respected. However
it does not correspond to the relativistic case either. To be kept
relativistic, we need a mass $10000$ times smaller or less ($\sim
2 \times 10^{-5}$ [eV]).\\

In Figure \ref{graf_vel_rel} the evolution of the velocity of the
neutrinos due to the expansion of the Universe is represented
graphically . We define the adimensional quantities $z =
\frac{M_{\nu}}{k_BT _{\nu}}$, $y =\frac{v _{\nu}}{c}$. It is
indicated in the graph that the time grows towards bigger values
in $z$. Clearly, we see that the neutrinos suffer a rapid
deceleration from the time of decoupling . Then, this deceleration
begins to diminish slowly, approaching a zero velocity.\\

All the estimations of $M _{\nu}$ indicate that we are in a zone
dominated by the Non-Relativistic approximation. The estimation
for the smallest masses ranges between $10^{-4}$ and $10^{-3}$
[eV]. Remembering that our top limit is $0.17$ [eV], we see that
we are currently in the region $0.6 < z <1012$, which is a very
wide range. Evaluating numerically in (\ref{veldesrel_norel}), we
obtain $v_{\nu} = 3.55 \times 10^{-3} c = 1065$ [$\frac{km}{s}$],
with a mass of $0.17$ [eV]. This velocity will be bigger if we use
smaller neutrino masses.\\

Up to now, we have assumed that the neutrinos are not affected by
the galactic potential, they are free particles and are not relics
from the Milky Way \cite{white1}-\cite{white2}. To check this
point, we consider the relation between kinetic and potential
energy of the neutrino in the Milky Way. That is:

$$\frac{m_{\nu}v^2}{2} = \frac{GMm_{\nu}}{R}$$
$$v = \sqrt{\frac{2GM}{R}}$$

Where $v$ would be the limit velocity where the potential energy
is comparable with the kinetic energy. Evaluating in $M \simeq 2
\times 10^{42}$ [kg] and $R \simeq 4.7 \times 10^{20}$ [m], mass
and radius of the Milky Way respectively, we obtain $v \simeq 754$
$\left [\frac{km}{s} \right]$. Since $v_{\nu} \gtrsim 1000$ $\left
[\frac{km}{s} \right]$, our supposition is correct.\\


\section{THE CNB DISTRIBUTION.}

To determine the effective neutrinos distribution (distribution
from Earth), we need to use equation (\ref{f_Rel}) in the comobile
system of the Universe. In addition to this, we will use the
photons distribution when they are decoupled, that is:

$$f(p,T) = \frac{1}{e^{\frac{pc}{kT}}-1}$$

That corresponds to an ultra-relativistic Bose-Einstein's
distribution with $RT = cte$. Currently, it has a temperature of
$2.73$ [K]. In addition, we saw in the previous chapter that a LIV
of the form:

$$E^2 = v_{max}^2p^2 + m^2c^4$$

Is not markedly different in its energy and velocity in comparison
to the usual Dispersion Relation ($v _{max} = c$). Because of this
the usual special relativity rules are valid. For instance,
Lorentz's Transformations for the coordinates of space-time and of
energy - momentum. Then, we can compare the neutrinos distribution
in the comobile system and Earth. It is possible to demonstrate
(See Appendix):

\begin{equation}
\label{f=f'} f'(p',T') = f(p,T)
\end{equation}

\begin{equation}
\label{E(E')} E = \gamma (E' - v_t p'\cos(\theta'))
\end{equation}

Where the primed elements refer to the reference system of Earth
and the non primed to the comobile. $\theta'$ is the angle that is
formed between the vision line and the direction of Earth motion
and $v_t$ is the Earth's velocity \cite{vel pec}. We can see that
the distribution function is invariant under Lorentz's
Transformation and the energy changes with the angle of vision.\\

Now we will analyze some cases. First, the photons of the CMB to
guide us because they are already very well known , and secondly,
the neutrinos. Using expression (\ref{E(E')}) we will determine
$p'$ as a function of $p$.\\

\subsection{Photons} In this case, we have
that $E = cp$, therefore the expression (\ref{E(E')}) is reduced
to:\\

$$p = \frac{1 - \frac{v_t}{c}\cos(\theta')}{\sqrt{1-\left(\frac{v_t}{c}\right)^2}}p'$$

Replacing in (\ref{f=f'}), we obtain:

$$f'(p',T'_{\gamma}) = f\left(\frac{1-\frac{v_t}{c}\cos(\theta')}{\sqrt{1-\left(\frac{v_t}{c}\right)^2}}p',T_{\gamma}\right)$$

As the photons, after being decoupled, continue with a
distribution of the form:

$$f_{\gamma} = \frac{1}{e^{\frac{pc}{k_BT_{\gamma}}}-1}$$

We can leave our expression as:

$$f'(p',T'_{\gamma}) = f\left(p',T_{\gamma}\frac{\sqrt{1-\left(\frac{v_t}{c}\right)^2}}{1
- \frac{v_t}{c}\cos(\theta')}\right)$$

Therefore, the photons distribution detected from Earth, $f'$, in
a specific direction, will be of the same form that the one detected in
the comobile system to the Universe, but with a different
temperature given by:

$$T'_{\gamma} = T_{\gamma}\frac{\sqrt{1-\left(\frac{v_t}{c}\right)^2}}{1
- \frac{v_t}{c}\cos(\theta')}$$

If we consider that $v_t \ll c$, we have:

$$T'_{\gamma} \simeq T_{\gamma}\left(1 +
\frac{v_t}{c}\cos(\theta')\right)$$

\begin{equation}
\label{Dipol} \frac{\Delta T_{\gamma}}{T_{\gamma}} \simeq
\frac{v_t}{c}\cos(\theta')
\end{equation}

that is known as the Dipolar Moment, and is of the order of
$10^{-4}$.\\

\subsection{Neutrinos} Now, we have particles
with mass. Currently, the neutrinos are Non-Relativistic,
therefore $E = m_{\nu} c^2 + \frac{p^2}{2m_{\nu}}$ for both the
comobile and Earth systems. Evaluating in (\ref{E(E')}) and using
the approximation $v_t \ll c$ up to second order in $p' $ and
$v_t$, we obtain:

$$p^2 = p'^2 - 2m_{\nu}v_tp'\cos(\theta') + m_{\nu}^2v_t^2$$

Evaluating in (\ref{f=f'}), we have:

\begin{equation}
\label{f'_neutrin} f'(p',T'_{\nu}) = f\left(\sqrt{p'^2 -
2m_{\nu}v_tp'\cos(\theta') + m_{\nu}^2v_t^2},T_{\nu}\right)
\end{equation}

In this case is impossible to find a relation between $T'_{\nu}$
and $T_{\nu}$, but we know that the distribution is given by
(\ref{f_Rel}). Seemingly, we can only notice the effects
graphically. To facilitate our analysis, it will be helpful to
define the number of neutrinos per solid angle $d\Omega'$ of
momentum as:

$$\frac{dN}{d\Omega'} = \frac{gV}{(2\pi
\hbar)^3}f'(p',T'_{\nu})p'^2dp$$

With this, we can obtain the distribution function of the number
of particles:

\begin{equation}
\label{F'} F'(p',T'_{\nu}) = \frac{gV}{(2\pi
\hbar)^3}f'(p',T'_{\nu})p'^2
\end{equation}

In our case, the distribution function $F'$ will be:

$$F'(p',T'_{\nu}) \propto \frac{p'^2}{e^{\frac{\sqrt{p'^2 -
2m_{\nu}v_tp'\cos(\theta') + m_{\nu}^2v_t^2}c}{k_BT_{\nu}}}+1}$$

\subsection{ANALYSIS}

To do our analysis, it is useful to introduce the adimensional
variables $x = \frac{p'c}{k_BT_{\nu}}$, $a(\theta') =
\frac{m_{\nu} v_tc}{k_BT_{\nu}} \cos (\theta') $ and $b = a(0)$.
With these parameters, our distribution is:

\begin{equation}
\label{F_rel}F' \propto \frac{x^2}{e^{\sqrt{x^2 - 2ax + b^2}}+1}
\end{equation}

Considering a terrestrial velocity $v_t \simeq 300$
[$\frac{km}{s}$], we can see that $b \simeq 1$ for $M_{\nu} =
0.17$. It means that $-1 \leq a(\theta') \leq 1$. This range will
be smaller if we use a smaller mass, but then the Non-Relativistic
approximation is less precise.\\

\begin{figure}[htb]
    \begin{center}
        \includegraphics[width=0.55\textwidth]{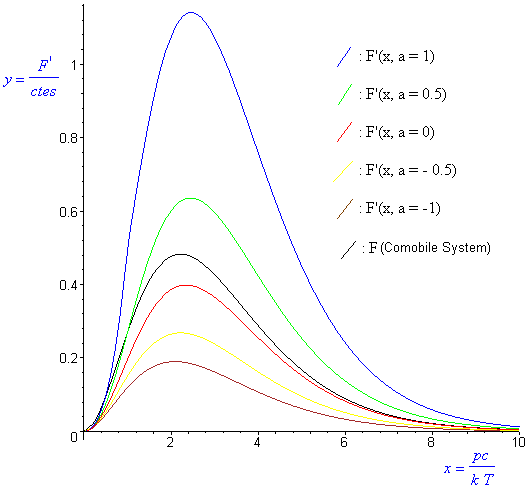}
    \end{center}
    \caption[t\'tulo indice (corto)]{{\footnotesize Primordial Neutrinos Distribution in the current age (Ec \ref{F_rel}) for different values of $a$ and $b \simeq 1$ that corresponds to $M_{\nu} = 0.17$ [eV]. The black curve represents to the distribution
    in the comobile system.}}
    \label{grafF-rel}
\end{figure}

In Figure \ref{grafF-rel} it is shown $F'$; here we have used a
value of $b\sim 1$ and some representative values of $a
(\theta') $ (See Table \ref{direc}).\\

\begin{table}
\begin{center}
\begin{tabular}{|c|c|c|} \hline
\textbf{$a(\theta')$} & \textbf{Direction of Observation}\\
\hline \hline
         $1$           &    In favour of the Terrestrial Movement      \\
        $0.5$          & $60^0$ deflected to the Terrestrial Movement  \\
         $0$           &  Perpendicular to the Terrestrial Movement   \\
       $-0.5$          & $120^0$ deflected to the Terrestrial Movement \\
        $-1$           &   Against the Terrestrial Movement     \\ \hline
\end{tabular}
\caption{Directions of Observation.}{{\footnotesize Values of $a
(\theta') $ used in Figure \ref{grafF-rel} with the corresponding
direction of observation.}} \label{direc}
\end{center}
\end{table}

Let's remind that the distribution $F'$ represents the particles
number that come from certain direction and momentum. We see in
Figure \ref{grafF-rel} that the distribution suffers a loss of
homogeneity, which is translated in more neutrinos observed in
favour of the Earth's movement, but simultaneously the form of the
distribution function is altered much more with regard to the
distribution of the comobile system. If we move away from this
direction, the neutrino number detected diminishes considerably
and the small momentum are favored.\\

The distribution maximum must fulfill the  equation:

\begin{equation}
\label{n_eq}\left(2\sqrt{x_{max}^2 - 2ax_{max} + b^2} - x_{max}^2
+ ax_{max}\right)e^{\sqrt{x_{max}^2 - 2ax_{max} + b^2}} +
2\sqrt{x_{max}^2 - 2ax_{max} + b^2}=0
\end{equation}

It is complicated to find a general expression for $x_{Max}$, although a numerical
treatment is readily available. As an example, 
we can study the extreme cases $a (\theta) = b$ and $a
(\theta) =-b$ where $b \simeq 1$ for $M _ {\nu} = 0.17$ [eV].
Evaluating in (\ref {n_eq}), we obtain:

$$x_{max}(a=b) = 2.463$$
$$x_{max}(a=-b) = 2.091$$

This means that the momentum of the majority of detected neutrinos
will be:

$$2.091 \leq \frac{p'c}{k_BT_{\nu}} \leq 2.463$$

This will be a useful information to plan the detectors.\\

\begin{figure}[htb]
    \begin{center}
        \includegraphics[width=0.55\textwidth]{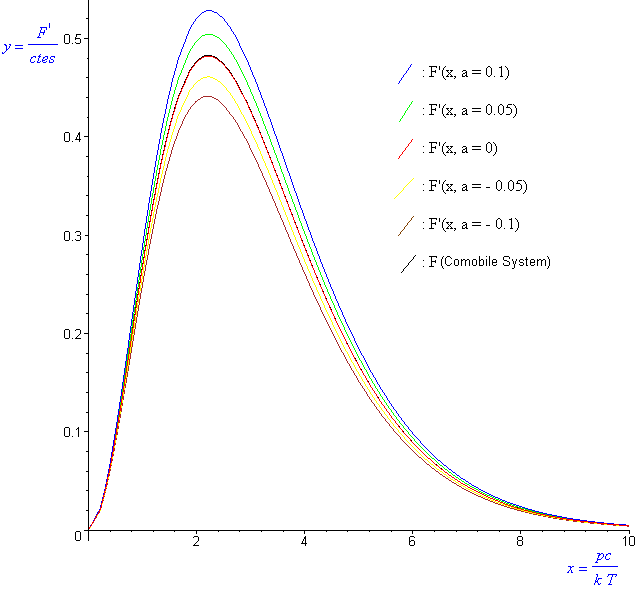}
    \end{center}
    \caption[t\'tulo indice (corto)]{{\footnotesize Primordial Neutrinos Distribution in the current age (Ec \ref{F_rel}) for different values of $a$ and $b \simeq 0.1$ that corresponds to $M_{\nu} = 0.017$ [eV]}}
    \label{grafF-rel_10}
\end{figure}

\begin{figure}[htb]
    \begin{center}
        \includegraphics[width=0.55\textwidth]{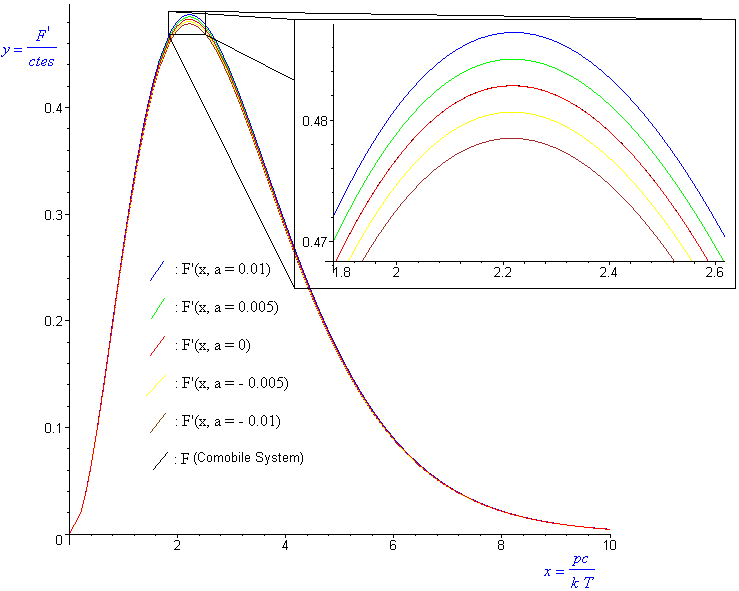}
    \end{center}
    \caption[t\'tulo indice (corto)]{{\footnotesize Primordial Neutrinos Distribution in the current age (Ec \ref{F_rel}) for different values of $a$ and $b \simeq 0.01$ that corresponds to $M_{\nu} = 0.0017$ [eV]}}
    \label{grafF-rel_100}
\end{figure}

\begin{figure}[htb]
    \begin{center}
        \includegraphics[width=0.55\textwidth]{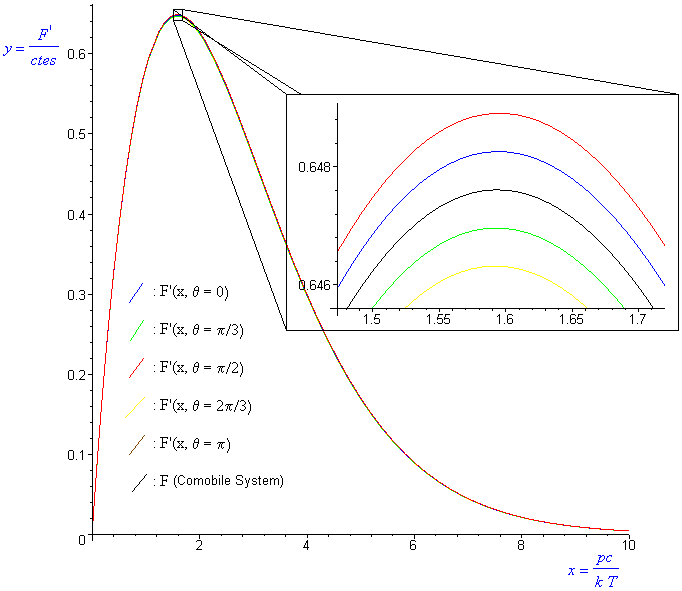}
    \end{center}
    \caption[t\'tulo indice (corto)]{{\footnotesize Primordial Photons Distribution in the current age (Ec \ref{F_rel}) for different values of the observation angle $\theta'$.}}
    \label{grafF-fot}
\end{figure}

Now, if we use a smaller mass, the differences between different
directions in the distributions diminish. In Figures
\ref{grafF-rel_10} and \ref {grafF-rel_100} we can see the
distributions with a mass $10$ and $100$ times smaller, where the
Non-Relativistic approximation can be still valid. To compare, the
photons distribution appears in Figure \ref{grafF-fot} for
different observation angles. Comparing this with Figure \ref
{grafF-rel_100}, we see that the effect produced in the neutrinos
distribution is bigger always than the produced in the photons.\\

This happens because the photons always go to a bigger velocity
than the terrestrial ($c \gg v_t$), therefore Earth would be
almost still with regard to the comobile system, doing that the
isotropy almost does not change. On the other hand, if the
neutrinos acquire mass, they will be found submitted to a
deceleration as the Universe expands (See Figure
\ref{graf_vel_rel}) so that  that currently the neutrinos are
Non-Relativistic, with a velocity not much higher than
$v_t$. This means that the effect of the terrestrial movement
begins being important in the velocities addition and will be
increased with the time due to the constant cooling of the
neutrinos. This will  reach the point in which the
neutrino velocity will be much smaller than $v_t$ and, practically,
the planetary movement will predominate. This will be reflected in
an increase of the distribution in the direction of the
terrestrial movement.\\

\section*{CONCLUSION.}

\begin{figure}[htb]
    \begin{center}
        \includegraphics[width=0.8\textwidth]{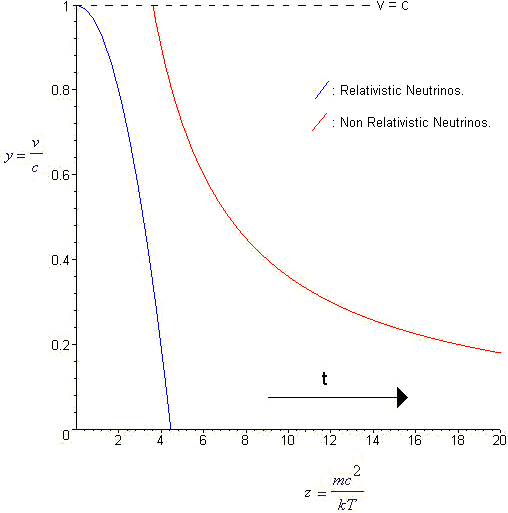}
    \end{center}
    \caption[Vel. Neutrinos Desacoplados Relativista]{{\footnotesize Neutrino Velocity Representation (Ecs. \ref{veldesrel_rel} y \ref{veldesrel_norel}). it is had being dominated by the relativistic expression (Blue) and then for the Non-Relativistic (Red). The general expression would be a composition of both.}}
    \label{graf_vel_rel}
\end{figure}

The mass of the neutrinos brought important modifications to its
velocity. Without mass, the neutrinos would have supported a
constant velocity and equal to the light velocity, $c$. On the
other hand, with non zero masses, its velocity is affected by a strong
deceleration (See Figures \ref{graf_vel_rel}), therefore they are
Non-Relativistic nowadays. As we have developed an expression for
the velocity with regard to the comobile system to the expansion
of the Universe, it is necessary to use the addition of velocities
to determine the mean neutrino velocity relative to Earth. Thus,
we use Lorentz's Transformations since the LIV did not bring any
important effect. The difference that is produced in its velocity
with and without LIV is of $\sim 10^{-20}$ \%, which is totally
negligible. Then, we can use the invariance of the distribution
function to relate the comobile system to the terrestrial.\\

In the same way, the mass of the neutrinos brought important
changes to the distribution. Unlike the photons, it was not
possible to introduce a similar term to the Dipolar Moment because
the temperature would depend on $p'$. Greater the mass greater the
effect. In addition, the distribution is widely favored in the
Earth's direction, but if we move away from this direction, the
neutrinos number diminishes. In spite of that the variation
depends greatly on the mass; as time goes by, the neutrinos will
be cooling diminishing little by little its velocity. This means
that in some moment the velocity of the neutrinos will be less than
the terrestrial speed. In the future, the neutrinos will be almost still
in comparison to the Earth's velocity. In this moment, we will
only detect the neutrino that "crash" with Earth when it
advances.\\

To sum up, we see that the existence of the neutrino mass produces
a relatively important effect in its evolution, which is reflected
in the perception that we have of them especially in the loss of
homogeneity in the distribution function. Thus, for its detection
is advisable to use detectors of neutrinos directed in favour to
the terrestrial movement or to use a satellite located in someone
of Lagrange's points of the Solar System to keep the isotropic
distribution of the comobile system, as the satellite \textbf
{Planck Surveyor} that will observe \textbf{CMB} \cite{planck}.\\

\section*{Acknowledgments}
The authors want to thank A. Reisenegger for an interesting discussion.
The work of JA and PG was partially supported by Fondecyt \# 1060646.


\section*{APPENDIX.}

The special relativity rules say us that two reference systems can
be related by:

$$\vec{x} = \vec{x'}_{\perp} + \gamma (\vec{x'}_{\parallel} + \vec{v}_t t')$$
$$t = \gamma \left(t' + \frac{\vec{v}_t\cdot\vec{x'}}{c^2}\right)$$

and

$$\vec{p} = \vec{p'}_{\perp} + \gamma \left(\vec{p'}_{\parallel} + \frac{\vec{v}_t}{c^2} E'\right)$$
$$E = \gamma (E' + \vec{v}_t\cdot\vec{p'})$$

Where the primed reference system is moving away from the non
primed to a velocity $\vec{v}$. The coefficients with the
subscripts $\parallel$ and $\perp$ represent the parallel and
perpendicular components of the velocity $\vec{v}$ respectively.
Since we are considering a particle in the universe, our primed
and non primed reference systems will be, respectively, Earth and
comobile system to the Universe expansion, therefore $\vec{v}$ is
the planet velocity. Thus, from now, we will call its
$\vec{v}_t$.\\

\begin{figure}[htb]
   \begin{center}
        \includegraphics[width=0.8\textwidth]{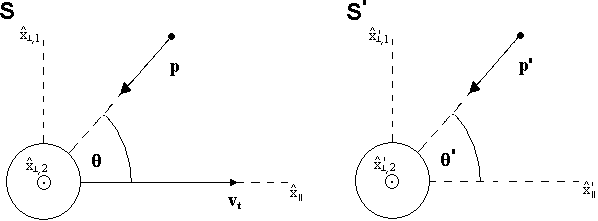}
    \end{center}
    \caption[t\'itulo indice (corto)]{{\footnotesize Description of both reference systems. $S$: Comobile reference system to the Universe expansion. Earth has a velocity $v_t$ and the neutrino has momentum $p$. Between both we have the vision angle $\theta$. $S'$: Earth reference system. Earth is still and the neutrino has momentum $p'$. We have the vision angle $\theta' $ measured from Earth. The coordinates system of $S$ and $S'$ are related by the Lorentz's Transformation.}}
    \label{sist_ref}
\end{figure}

If particles go to Earth along the vision line (See Figure \ref
{sist_ref}), the Lorentz's transformation can be written as:

\begin{equation}
\label{Trans_Lor_esp} x_{\perp,i} = x'_{\perp,i}~~~~x_{\parallel}
= \gamma(x'_{\parallel} + v_t t')~~~~t = \gamma \left(t' +
\frac{v_t x'_{\parallel}}{c^2}\right)
\end{equation}

\begin{equation}
\label{Trans_Lor_mom} p_{\perp,i} = p'_{\perp,i}~~~~p_{\parallel}
= \gamma \left(p'_{\parallel} - \frac{v_t}{c^2}E'\right)~~~~E =
\gamma(E' - v_t p'_{\parallel})
\end{equation}

Where $i=1\textrm{, } 2$ label both perpendicular coordinates to
$\vec{v}_t$. Its differential form considering an instantaneous
measurement from Earth, is $t' =cte$ or $dt' =0$, giving:

\begin{equation}
\label{Trans_Lor_difesp} dx_{\perp,i} =
dx'_{\perp,i}~~~~dx_{\parallel} = \gamma dx'_{\parallel}~~~~dt =
\gamma \frac{v_t dx'_{\parallel}}{c^2}
\end{equation}

\begin{equation}
\label{Trans_Lor_difmom} dp_{\perp,i} =
dp'_{\perp,i}~~~~dp_{\parallel} = \gamma \left(dp'_{\parallel} -
\frac{v_t}{c^2}dE'\right)~~~~dE = \gamma(dE' - v_t
dp'_{\parallel})
\end{equation}

Now, when we count the particles number from Earth in a specific
direction instantaneously ($dt' =0$), inside a volume $d^3r'$ we
have $dN$ particles with momentum between $\vec{p'}$ and $\vec{p'}
+ d\vec{p'}$. In addition, we know that $dN$ is given by:

\begin{equation}
\label{dNf'} dN = f'(p',T')d^3p'd^3r'
\end{equation}

Where $f' $ is the distribution function on Earth. In the comobile
system, the particles are in a volume $d^3r$ and with values of
momentum between $\vec{p}$ and $\vec{p} + d\vec{p}$, but in a $dt$
time, given by (\ref{Trans_Lor_difesp}) (different from zero
because $dt' =0$) some particles enter or exit of $d^3r$. Thus,
the particle number, in this system, is given by:

\begin{equation}
\label{dNf} dN = f(p,T)d^3pd^3r + f(p,T)d^3p d\vec{S} \cdot
\vec{u} dt
\end{equation}

Where $\vec{u} = c^2\frac{\vec{p}}{E}$ is the particle's velocity
and $d\vec{S}$ is the differential area, with normal direction.
Both expressions for $dN$ are, simply, a variation of the
continuity equation:

$$\frac{\partial \rho}{\partial t} + \nabla \cdot (\rho \vec{u}) = 0$$

With $\rho = f(p, T) d^3p$. Since $dN$ must be the same in both
systems, we must equal (\ref{dNf'}) and (\ref{dNf}). Then:

\begin{equation}
\label{relacion_f}f'(p',T')d^3p'd^3r' = f(p,T)d^3p\left(d^3r +
c^2dt\frac{\vec{p} \cdot d\vec{S}}{E}\right)
\end{equation}

\begin{figure}[htb]
    \begin{center}
        \includegraphics[width=0.8\textwidth]{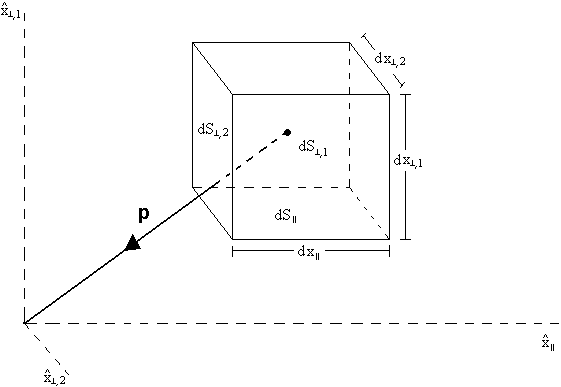}
    \end{center}
    \caption[t\'itulo indice (corto)]{{\footnotesize Representation of the volume element $d^3r$, where we see the surface elements. We can see, clearly, the vectorial direction of $\vec{p}$ and $d\vec{S}$.}}
    \label{d3r}
\end{figure}

With (See Figure \ref{d3r}):

$$d^3r = dx_{\parallel} \wedge dx_{\perp,1} \wedge dx_{\perp,2}$$
$$d\vec{S} = -(dx_{\parallel} \wedge dx_{\perp,1}\hat{x}_{\perp,2} + dx_{\parallel} \wedge dx_{\perp,2}\hat{x}_{\perp,1} + dx_{\perp,1} \wedge dx_{\perp,2}\hat{x}_{\parallel})$$
$$\vec{p} = -(p_{\parallel}\hat{x}_{\parallel} + p_{\perp,1}\hat{x}_{\perp,1} +
p_{\perp,2}\hat{x}_{\perp,2})$$

Where $ \wedge $ represents the anti-commutative product between
the differentials. Evaluating in (\ref{relacion_f}), using (\ref
{Trans_Lor_mom}) and (\ref{Trans_Lor_difesp}), we have:

\begin{equation}
\label{rel_f_f'} f'(p',T')d^3p'd^3r' = f(p,T)d^3pd^3r'\frac{E'}{E}
\end{equation}

Replacing (\ref{Trans_Lor_difmom}) in $d^3p = dp_{\perp,1} \wedge
dp_{\perp,2} \wedge dp_{\parallel}$, we obtain:

$$d^3p = dp'_{\perp,1} \wedge dp'_{\perp,2} \wedge \gamma\left(dp'_{\parallel} - \frac{v_t}{c^2} dE'\right)$$

But we know that ${E'}^2 = c^2({p'}_{\perp,1}^2 +{p'}_{\perp,2}^2
+{p'}_{\parallel}^2) + m^2c^4$. Deriving, we obtain the relation
$E'dE' = c^2(p'_{\perp,1}dp'_{\perp,1} + p'_{\perp,2}dp'_{\perp,2}
+ p'_{\parallel}dp'_{\parallel})$. Evaluating:

$$d^3p = dp'_{\perp,1} \wedge dp'_{\perp,2} \wedge dp'_{\parallel}\gamma\left(1 - \frac{p'_{\parallel}}{E'}v_t\right)$$

Where we have used that $dp'_{\perp, i}  \wedge dp'_{\perp, i} =
0$ for anti-conmutativity. Using (\ref{Trans_Lor_mom}), $d^3p$
stays:

$$d^3p = d^3p'\frac{E}{E'}$$

Then, (\ref{rel_f_f'}) is reduced to:

\begin{equation}
f'(p',T') = f(p,T)
\end{equation}

This means that the distribution function is Lorentz invariant. In reference \cite{f_inv}, this has been discussed differently.
They used:

$$dN = f(p,T)d^3pd^3r$$

Naturally, they obtained that $f$ is not Lorentz invariant.\\


\end{document}